\documentclass[aps,prd,superscriptaddress,amsfont,graphicx,nofootinbib,preprintnumbers]{revtex4}%
\usepackage{color,graphicx,epsfig}
\usepackage{ifpdf}
\usepackage{amsmath}
\usepackage{bm}
\usepackage{color}
\usepackage[english]{babel}
\usepackage{graphicx}%
\usepackage{amsfonts}%
\usepackage{amssymb}
\usepackage{braket}
\usepackage{hyperref}

\definecolor{nicered}{rgb}{0.7,0.1,0.1}
\definecolor{nicegreen}{rgb}{0.1,0.5,0.1}
\hypersetup{colorlinks,citecolor= nicegreen,linkcolor= nicered}

\newenvironment{Eqnarray}{\arraycolsep 0.14em\begin{eqnarray}}{\end{eqnarray}}
\def\beqa{\begin{Eqnarray}}
\def\eeqa{\end{Eqnarray}}
\newcommand{\no}{\nonumber}

\newcommand{\beq}{\begin{equation}}
\newcommand{\eeq}{\end{equation}}
\newcommand{\bea}{\begin{eqnarray}}
\newcommand{\eea}{\end{eqnarray}}

\def\lsim{\mathrel{\rlap{\lower4pt\hbox{\hskip1pt$\sim$}}
     \raise1pt\hbox{$<$}}}         
\def\gsim{\mathrel{\rlap{\lower4pt\hbox{\hskip1pt$\sim$}}
     \raise1pt\hbox{$>$}}}         

\begin{document}

\title{The Higgs program and open questions in particle physics and cosmology}

\author{Beate Heinemann$^{1,2a}$, and Yosef Nir$^{3b}$}
\affiliation{$^1$Deutsches Elektronen-Synchrotron, 22607 Hamburg, Germany\\
$^2$Albert-Ludwigs-Universit{\"a}t Freiburg, Physikalisches Institut, 79104 Freiburg, Germany\\
$^3$Department of Particle Physics and Astrophysics, Weizmann Institute of Science, Rehovot, Israel 7610001}
\email{$^a$beate.heinemann@desy.de; $^b$yosef.nir@weizmann.ac.il}

\begin{abstract}
\noindent
The Higgs program is relevant to many of the open fundamental questions in particle physics and in cosmology. Thus, when discussing future collider experiments, one way of comparing them is by assessing their potential contributions to progress on these questions. We discuss in detail the capabilities of the various proposed experiments in searching for singlet scalars, which are relevant to several of the open questions, and in measuring Higgs decays to fermion pairs, which are relevant to the flavor puzzles. On other interesting questions, we list the most relevant observables within the Higgs program.
\end{abstract}

\maketitle

\section{Introduction}
The jewel in the crown of the achievements of the LHC experiments to date is the Higgs boson discovery in July 2012 \cite{Aad:2012tfa,Chatrchyan:2012xdj}. The discovery of a Higgs boson is, however, not just an end of a story - a quest that began with the theoretical predictions of Brout, Englert \cite{Englert:1964et} and Higgs \cite{Higgs:1964pj} - but also a beginning of one.
While the presence of the Higgs boson provides a solution to the question of how fundamental particles can acquire mass, it does not actually explain the mass values themselves, and it also raises new questions. ``The Higgs program" is a major research project at the LHC, and at all proposed future collider experiments. From the experimental side, the Higgs program refers to a large set of measurements aimed to learn the detailed properties of this unique particle. From the theoretical side, this is also a very exciting program, as it touches upon several open questions and puzzles in particle physics and particle cosmology.

In this article, we explore the relationship of the properties of the Higgs boson to big questions in the field of particle physics and cosmology. This work was initiated by discussions in the Scientific Policy Committee of CERN in the context of discussions on future accelerators. All proposed future accelerators consider measurements of the Higgs boson a major part of their scientific program, and this short article is designed to summarize concisely the conclusions from the currently available literature that relates Higgs precision measurements to fundamental open questions about our Universe. We also summarize the precision these future colliders estimate, and try to contrast that with the precision required to answer a given question. It is worth noting that for many of the questions there are also important observables unrelated to the Higgs boson measurements (at both collider and non-collider experiments), but discussing those is beyond the scope of this article.

Here is a list of seven open questions at the forefront of particle physics and cosmology, to which the Higgs program may provide answers:
\begin{enumerate}
\item Is $h$ the only scalar degree of freedom?
\item Is $h$ elementary?
\item What keeps $m_h^2\ll m_{\rm Pl}^2$?
\item Was the electroweak phase transition first order?
\item Did CP violating $h$ interactions generate the baryon asymmetry?
\item Are there light SM-singlet degrees of freedom (in particular, related to Dark Matter)?
\item What is the solution of the flavor puzzle(s)?
\end{enumerate}

In what follows, we focus on two topics. In Section \ref{sec:singlet}, we describe the search for singlet scalars. We choose this topic for four reasons:
\begin{itemize}
\item It is relevant to a large number (four) of the above questions;
\item The Higgs program is likely to be the unique portal to such particles;
\item The search for singlet scalars by itself constitutes a rich and broad experimental program;
\item For the question of the electroweak phase transition, there is a clear target for the required accuracy that will give a definite answer on this question.
\end{itemize}
In Section \ref{sec:flavor}, we discuss the measurements that are relevant to the flavor puzzles. Again, there are several reasons for this choice of topic:
\begin{itemize}
\item The flavor puzzles are long standing, and the Higgs program provides an opportunity to measure new flavor parameters and make progress;
\item The main ingredient in the flavor-related Higgs program -- the measurement of Yukawa couplings -- is also relevant to other questions.
\item The flavor related measurements by themselves  constitute a rich and broad experimental program;
\item In contrast to the question of electroweak phase transition, for the flavor measurements there is no lower bound on the size of new physics effects. Instead, the better accuracy, the higher the scale of new physics to which there is sensitivity.
\end{itemize}
All other topics are discussed briefly in Section \ref{sec:addque}, where we describe the most relevant experimental measurements to each physics question, but do not compare the capabilities of the various experiments (which is still work in progress). In Section \ref{sec:con} we present three important conclusions about the scientific significance of the Higgs program.

\section{Experiments}
\label{sec:exp}
For the purpose of discussing the expected accuracy of future measurements, the following future collider parameters (integrated luminosity ${\cal L}$, center of mass energy $\sqrt{s}$, and, where relevant, polarization) are assumed:
\begin{itemize}
\item HL-LHC: $pp$ collider, ${\cal L}=3-4$~ab$^{-1}$ at $\sqrt{s}=14$~TeV (2026 to late 2030s: $\gsim 10$ years)~\cite{Bordry:2018gri}.
\item ILC: $e^+e^-$ collider, with $e^-$ 80\% polarized and $e^+$ 30\% polarized, to be operated in two stages (22 years total including shutdowns)~\cite{Bambade:2019fyw}:
\begin{enumerate}
\item ILC250: ${\cal L}=2$~ab$^{-1}$ at $\sqrt{s}=250$~GeV (11 years),
\item ILC500: ${\cal L}=4$~ab$^{-1}$ at $\sqrt{s}=500$~GeV (10 years). This includes also a 1-year run at $\sqrt{s}=365$~GeV.
\end{enumerate}
\item CLIC: $e^+e^-$ collider, with $e^-$ 80\% polarized and $e^+$ not polarized, to be operated in  three stages (26 years including shutdowns)~\cite{Bordry:2018gri}:
\begin{enumerate}
\item CLIC380: ${\cal L}=1.0$~ab$^{-1}$ at $\sqrt{s}=380$~GeV (8 years),
\item CLIC1500: ${\cal L}=2.5$~ab$^{-1}$ at $\sqrt{s}=1400$~GeV (7 years),
\item CLIC3000: ${\cal L}=5.0$~ab$^{-1}$ at $\sqrt{s}=3000$~GeV (7 years).
\end{enumerate}
\item CEPC: $e^+e^-$ collider, without polarization, ${\cal L}=5.6$~ab$^{-1}$ at $\sqrt{s}=240$~GeV (7 years)~\cite{CEPCStudyGroup:2018ghi}. It is assumed that there are two experiments collecting the data, and the datasets are combined.
\item FCC-ee: $e^+e^-$ collider, without polarization, to be operated in two stages~\cite{Bordry:2018gri}:
\begin{enumerate}
\item FCC240: ${\cal L}=5.0$~ab$^{-1}$ at $\sqrt{s}=240$~GeV (3 years).
\item FCC365: ${\cal L}=1.5$~ab$^{-1}$ at $\sqrt{s}\approx 365$~GeV (5 years).
\end{enumerate}
In both cases it is assumed that there are two experiments collecting the data, and the datasets are combined.
\item LHeC: $ep$ collider, colliding the 7~TeV HL-LHC protons with 60~GeV electrons, aiming to deliver ${\cal L}=1$~ab$^{-1}$ within 12 years. For the last 4 years the HL-LHC is assumed not to operate concurrently for $pp$ collisions~\cite{Bordry:2018gri}.
\item HE-LHC:  $pp$ collider, ${\cal L}=10$~ab$^{-1}$ at $\sqrt{s}=27$~TeV (20 years)~\cite{Bordry:2018gri}. For the numbers presented in Table \ref{tab:yff}, ${\cal L}=15$~ab$^{-1}$ is considered, corresponding to two collider experiments combining at least 75\% of the data~\cite{Mangano:2018mur}.
\item FCC-hh:  $pp$ collider, ${\cal L}=20$~ab$^{-1}$ at $\sqrt{s}=100$~TeV (25 years)~\cite{Bordry:2018gri}. For the numbers presented in Table \ref{tab:yff}, ${\cal L}=30$~ab$^{-1}$ is considered, corresponding to two collider experiments combining at least 75\% of the data~\cite{Mangano:2018mur}.
\item Muon collider: a muon collider is also a very interesting option to collide leptons at very high energies. This is, however, not considered in this report as the studies related to this are currently less advanced.
\end{itemize}

\section{Singlet scalars}
\label{sec:singlet}
One of the experimentally most challenging extensions of the Standard Model (SM), yet one that is relevant to many interesting Higgs-related questions, is that of a singlet scalar. In fact, it appears with relation to four of the seven open questions that we posed:
\begin{enumerate}
\item Is $h$ alone?
\item What keeps $m_h^2\ll m_{\rm Pl}^2$?
\item Was the electroweak phase transition first order?
\item Are there light SM-singlet degrees of freedom?
\end{enumerate}

The simplest extension is to add to the SM a real scalar field $S(x)$, in the $(1,1)_0$ (gauge-singlet) representation. The most general renormalizable Lagrangian is \cite{Huang:2016cjm,Craig:2013xia,Curtin:2014jma} (we use the notations of \cite{Huang:2016cjm})
\beq
{\cal L}={\cal L}_{\rm SM}+\frac12(\partial_\mu S)(\partial^\mu S)-\frac{m_s^2}{2}S^2-\frac{a_s}{3}S^3-\frac{\lambda_s}{4}S^4
-\lambda_{hs}\Phi^\dagger\Phi S^2-2a_{hs}\Phi^\dagger\Phi S,
\eeq
where $\Phi$ is the Higgs doublet in the $(1,2)_{+1/2}$ representation. In the general case, both $\Phi$ and $S$ can acquire VEVs:
\beq
\langle\Phi\rangle=(0,v/\sqrt2),\ \ \ \langle S\rangle=v_s,
\eeq
and the Higgs and singlet fields mix:
\beq\label{eq:hsmix}
\sin2\theta=\frac{4v(a_{hs}+\lambda_{hs}v_s)}{M_h^2-M_s^2},
\eeq
where $M_h$ and $M_s$ are the mass eigenvalues of the two scalar mass eigenstates ($M_h\simeq125$ GeV).

Deviations from the SM predictions for various couplings are often parameterized by ``$\kappa$ parameters":
\beq
\kappa_x\equiv\frac{g_{hxx}}{g_{hxx}^{\rm SM}}.
\eeq
Within the SM extended by singlet scalars, the couplings of the Higgs boson to all fermion and vector boson pairs are modified by a universal factor. (For a discussion, see Section I of Ref. \cite{Craig:2013xia}.)  Defining $\delta g_h$ via
\beq
\kappa_x\equiv1+\delta g_h,
\eeq
we have \cite{Huang:2016cjm}
\beq\label{eq:delzh}
\delta g_h\approx(\cos\theta-1)-\frac{|a_{hs}+\lambda_{hs}v_s|^2}{8\pi^2}I_B(M_h^2;M_h^2,M_s^2)
-\frac{|\lambda_{hs}|^2v^2}{16\pi^2}I_B(M_h^2;M_s^2,M_s^2),
\eeq
where \cite{Fan:2014axa}
\beq
I_B(p^2;m_1^2,m_2^2)=\int_0^1dx\frac{x(1-x)}{x(1-x)p^2-xm_1^2-(1-x)m_2^2}.
\eeq
%
%
%

As concerns Eq. (\ref{eq:delzh}), we emphasize the following points:
\begin{itemize}
\item The first term on the right hand side (RHS) is a consequence of
  the mixing  (see Eq. \ref{eq:hsmix}).
\item The second and third terms are a consequence of the wave function renormalization of the Higgs at one loop.
\end{itemize}

Usually, the first term dominates. This is not the case, however, in models where a $Z_2$ symmetry, under which $S$ ($\Phi$) is odd (even), is imposed. This scenario is considered the most difficult one for finding experimental evidence for first order electroweak phase transition~\cite{Curtin:2014jma}.
In this model, we have $a_{hs}=v_s=0$ and consequently $\theta=0$, so that the first two terms
of Eq.~(\ref{eq:delzh}) vanish. It follows that~\cite{Fan:2014axa}
\beq\label{eq:delzhztwo}
\delta g_h^{(Z_2)}=-\frac{|\lambda_{hs}|^2v^2}{16\pi^2}I_B(M_h^2;M_s^2,M_s^2)
=-\frac{|\lambda_{hs}|^2v^2}{16\pi^2 M_h^2}\left[1-\frac{4M_s^2{\rm arctan}\left(
\sqrt{\frac{M_h^2}{4M_s^2-M_h^2}}\right)}{\sqrt{M_h^2(4M_s^2-M_h^2)}}\right].
\eeq
In the limit that $M_s^2\gg M_h^2$ \cite{Fan:2014axa},
\beq
\delta g_h^{(Z_2)}\to +\frac{|\lambda_{hs}|^2v^2}{96\pi^2 M_s^2}.
\eeq
%

\subsection{$M_s<M_h/2$}
If $M_s<M_h/2$, then a new decay channel for the Higgs boson opens up, $h\to SS$, with
\beq
\Gamma(h\to SS)=\frac{\lambda_{hSS}^2}{8\pi M_h}\sqrt{1-\frac{4M_s^2}{M_h^2}},
\eeq
and
\beq
\lambda_{hSS}=+c_\theta^3\lambda_{hs}v
+c_\theta^2s_\theta (a_s+3\lambda_s v_s+2\lambda_{hs}v_s-2a_{hs})
+c_\theta s_\theta^2(3\lambda_{h}v-2\lambda_{hs}v)
+s_\theta^3(\lambda_{hs}v_s+a_{hs}).
\eeq

In the $Z_2$ limit
\beq
\lambda_{hSS}=\lambda_{hs}v,
\eeq
and, furthermore, $S$ is stable, and therefore, the $h\to SS$ mode contributes to $h\to{\rm invisible}$. Ref. \cite{Frugiuele:2018coc} collected the upper bounds and projections on BR($h\to$invisible). We quote updated results and future projections in Table~\ref{tab:inv}.

The BR($h\to$invisible) is determined directly by analyses searching
for a Higgs boson produced in association with a $Z$ boson (at lepton
and hadron colliders) or with forward jets (in VBF process at hadron
colliders), and the Higgs boson decaying invisibly which is observed
as missing transverse momentum in the experiments. Several $e^+e^-$
colliders are able to probe this with a precision of 0.3\%. The best
sensitivity is achieved by FCC$_{hh}$ with 0.025\%.

\begin{table}[t]
\caption{Current upper bounds and projections on ${\rm BR}_{\rm inv}\equiv{\rm BR}(h\to{\rm invisible})$ and ${\rm BR}_{\rm und}\equiv{\rm BR}(h\to{\rm undetected})$. All upper bounds are given at 95\% CL unless explicitly stated.
The LHC2 line updates the most recent bounds from CMS \cite{Sirunyan:2018owy} and ATLAS~\cite{Aaboud:2018sfi}.
Projections for future colliders are also shown: CLIC~\cite{Robson:2018zje}, CEPC~\cite{CEPCStudyGroup:2018ghi}, ILC~\cite{Barklow:2017suo} , FCC$_{\rm ee}$ and FCC$_{\rm hh}$~\cite{Mangano:2018mur}.
For FCC$_{\rm hh}$ the SM background from $H\to 4\nu$, corresponding to a BR of $0.113\%$, has been subtracted and the limit refers to additional invisible new physics contributions. For the LHC and HL-LHC BR$_{\rm und}$ is inferred assuming $\kappa_Z\leq 1$. The symbol "$-$" means that no value is available in the literature.
}
\label{tab:inv}
\begin{center}
\begin{tabular}{|c|c|c||c|c| } \hline\hline
\rule{0pt}{1.2em}%
Collider &  $\sqrt{s}$ [TeV] & ${\cal L}$ [ab$^{-1}$] & ${\rm BR}_{\rm inv}[\%]$ & ${\rm BR}_{\rm und}[\%]$ \\[2pt] \hline\hline
\rule{0pt}{1.2em}%
LHC1 & $7,8$ & $0.022$ & $37$ & 20 \\
LHC2 & $13$ & $0.036$ & $26$  & $-$  \\
LHC3 & $13$ & $0.300$ & $8.8\ (68\%)$ & $7.6\ (68\%)$ \\
HL-LHC & $14$ & $3$ & $2.6$ & $2.5$ \\

\hline\hline
CLIC & $0.38$ & $1$ & $0.69 (90\%)$ & $1.6$\\
CEPC & $0.25$ & $5.6$ & $0.3$  & $1.0$ \\
ILC & $0.25$ & $2$ & $0.3$ & $1.5$ \\
FCC$_{\rm ee}$ & $0.24$ & $5$ & $0.3$ & $0.87$  \\
FCC$_{\rm hh}$ & $100$ & $20$ & $0.025$ & $-$ \\
\hline\hline
\end{tabular}
\end{center}
\end{table}

In the case that no $Z_2$ symmetry is imposed and, in particular, $\sin\theta\neq0$, the mass eigenstate $S$ inherits the Higgs couplings, suppressed by $\sin\theta$:
\beq\label{eq:sff}
\lambda_{Sf\bar f}=\sin\theta\ y_f^{\rm SM},
\eeq
where $y_f^{\rm SM}$ is the corresponding SM Yukawa coupling. The $h\to SS$ decay will be followed by $S\to f\bar f$ where $m_f< M_s/2$ with coupling given by Eq. (\ref{eq:sff}). There are two complementary ways of searching for these decays. One is the direct search for the final four fermion state, which we discuss further below. The second is the indirect search, via a global fit to the Higgs couplings, which can reveal (or constrain) a contribution to the Higgs width that is not accounted for by the final states that are included in the fit \cite{Bechtle:2014ewa,Flacke:2016szy,Frugiuele:2018coc}, the so-called
\beq
h\to{\rm undetected}.
\eeq
Upper bounds on ${\rm BR}(h\to{\rm undetected})$ are given in Table
\ref{tab:inv}, and put an upper bound on $\sin^2\theta$ for a given
$M_s$. The branching ratios to ``undetected" final state are derived
from the observed cross sections and branching ratios using certain
constraints. For the LHC and HL-LHC this can only be done if $\kappa_Z \le 1.0$ is assumed, which holds for singlets with $M_s<M_h/2$ if the dominant effect is due to $\sin\theta\neq0$ (and in general only holds for a fraction of new physics models). For the $e^+e^-$ colliders it is possible to use a much more direct and very model-independent constraint ~\cite{Frugiuele:2018coc} ${\rm BR}(h\to {\rm undetected}) \le 1- (1-2\delta_\kappa)^2$, where $\delta_\kappa$ is the precision on the measurement of $\kappa_Z$ (CLIC380: 0.4\%, CEPC: 0.25\%, ILC: 0.38\%, FCCee: 0.22\%).


Examples of other relevant experimental searches are collected in Table \ref{tab:ssffff}. For the $\mu\mu bb$ final state it is expected that the HL-LHC~\cite{Atlas:2019qfx} will have a sensitivity down to ${\rm BR}(h\to \mu^+\mu^-b\bar b)\sim 5\times 10^{-5}$.  For the other decays no projections are available.
\begin{table}[t]
\caption{Upper bounds on $(\sigma_h/\sigma_h^{\rm SM})\times{\rm BR}(h\to SS\to \overline{f_1}f_1 \overline{f_2}f_2)$.}
\label{tab:ssffff}
\begin{center}
\begin{tabular}{|c|c|c|c|c|} \hline\hline
\rule{0pt}{1.2em}%
$\overline{f_1}f_1 \overline{f_2}f_2$ &  $M_s$ [GeV] & Upper bound & Ref.   \\[2pt] \hline\hline
\rule{0pt}{1.2em}%
$\tau\tau bb$ & $15-60$ & $0.03-0.12$ & CMS \cite{Sirunyan:2018pzn} \\
$\mu\mu bb$ & $25-63$ & $(1-8)\times 10^{-4}$ & CMS \cite{Khachatryan:2017mnf}, ATLAS \cite{Aaboud:2018esj} \\
$\tau\tau\tau\tau$ & $9-15$ & $0.2-0.3$ & CMS \cite{Khachatryan:2017mnf} \\
$\mu\mu\tau\tau$ & $25-63$ & $(1-5)\times 10^{-4}$ & CMS \cite{Khachatryan:2017mnf,Sirunyan:2018mbx} \\
$\gamma\gamma jj$ & $20-60$ & $0.034-0.173$ & ATLAS \cite{Aaboud:2018gmx} \\
\hline\hline
\end{tabular}
\end{center}
\end{table}

The total width of $h$ is, on one hand, suppressed by the suppression of the $h$-couplings to pairs of SM particles but, on the other hand, enhanced by the addition of the $h\to SS$ channel. It thus provides an interesting observable to probe this scenario.

The width of the $h$ boson can be inferred at the LHC using three methods:
\begin{itemize}
\item The cross section of off-shell $h$-boson production is proportional to $[(s-m_h^2)^2+m_h^2\Gamma_H^2]^{-1}$. Measuring it at different values of $s$ allows an extraction of the width under the assumption that the couplings of $h$ to the particles involved are independent of $s$ (this is not a good assumption for several BSM models). The most precise extraction~\cite{Sirunyan:2019twz} has been done in the $ZZ$ decay mode by the CMS collaboration using $\sim 80$~fb$^{-1}$ of run-2 data, and gives $3.2^{+2.8}_{-2.2}$ MeV. It is foreseen that, with the HL-LHC and improvements in the theoretical calculations, $\Gamma_H$ will be extracted with a precision of 20\% using this method~\cite{Cepeda:2019klc}.
\item Using the diphoton decay mode it is also possible to determine the width through interference effects. This is complementary, as well as theoretically robust, and will be able to probe values of $(8-22)\times\Gamma_h$~\cite{Cepeda:2019klc}.
\item Using the coupling measurements and assuming that $\kappa_Z\leq 1$ (which holds if the dominant effect is due to $\sin\theta\neq0$), $\Gamma_h$ can be extracted with a precision of 5\% with the HL-LHC. This determination of the width is only possible if one assumes that there are no undetected final states. We can either fix the width and determine the BR to undetected particles or we can set the latter to zero and determine the width.
\end{itemize}

In $e^+e^-$ colliders, $\Gamma_h$ can be measured based on $\sigma(e^+e^-\to Zh)$ and ${\rm BR}(h\to ZZ^*)$:
\beq
\frac{\sigma(e^+e^- \to Zh)}{{\rm BR}(h\to ZZ^*)}= \frac{\sigma(e^+e^- \to Zh)}{\Gamma(h\to ZZ^*)/\Gamma_h}= \left[\frac{\sigma(e^+e^- \to Zh)}{\Gamma(h\to ZZ^*)}\right]_{\rm SM}\times\Gamma_h,
\eeq
where we used the fact that, within the framework discussed here, both $\sigma(e^+e^- \to Zh)$ and $\Gamma(h\to ZZ^*)$ are modified from their SM values by the same factor, $\kappa_Z^2$.
The precision of this determination is then limited by the statistics of $h\to ZZ^*$ decays. In practice, for most colliders the $\kappa$-framework is used to extract the width parameter which benefits from additional measurements that are also sensitive to the width via similar reasoning.
In particular, both the statistical precision and the model dependence can be reduced significantly if also $\sigma(e^+e^- \to h\nu_e\nu_e)$ and BR($h\to WW^*$) as is possible for FCC$_{ee}$ at 365~GeV, CLIC (at all proposed energies) and ILC at 500~GeV. This explains the substantial improvement observed when higher energy data are included.
The estimated sensitivities of future $e^+e^-$ colliders to the total width are given in Table~\ref{tab:width}. It is much superior to the first and second determinations of the width at the LHC discussed above. A word of caution is, however, in place here. While the assumption that both the $Zh$ cross section and the $h\to ZZ$ decay width are modified by one simple factor, $\kappa_Z^2$, holds in the framework discussed in this subsection (extending the SM with a single light real singlet scalar), it is a model-dependent assumption, and cannot be assumed in general~\cite{Barklow:2017suo}. Thus, for ILC, the width has been extracted using an EFT fit with use of polarization and angular information instead. This is the reason that we quote also the method used in each experiment in Table~\ref{tab:width}.

\begin{table}
\caption{Precision on $\Gamma_h$ for the following $e^+e^-$ colliders: CLIC~\cite{Robson:2018zje}, CEPC~\cite{CEPCStudyGroup:2018ghi}, ILC~\cite{Bambade:2019fyw} and FCC$_{\rm ee}$~\cite{Mangano:2018mur}. Also shown is the method used to determine the width.}
\label{tab:width}
\begin{center}
\begin{tabular}{|c|c|c||c|c| } \hline\hline
\rule{0pt}{1.2em}%
Collider &  $\sqrt{s}$ [TeV] & ${\cal L}$ [ab$^{-1}$] & $\delta\Gamma_h/\Gamma_h$ [\%] & method \\[2pt]\hline\hline
\rule{0pt}{1.2em}%
CLIC 380 & $0.38$ & $1.0$ & $4.7$ & $\kappa$\\
CLIC 1.5 & $0.38+1.5$ & $2.5$ & $2.6$ & $\kappa$\\
CLIC 3.0 & $0.38+1.5+3$ & $5$ & $2.5$ & $\kappa$\\
ILC 250 & $0.25$ & $2.0$ & $2.4$  & EFT\\
ILC 500 & $0.25+0.5$ & $2.0+4.0$ & $1.6$ & EFT \\
CEPC & $0.25$ & $5.6$ & 2.8 & $\kappa$\\
FCC$_{\rm ee}$ 240 & $0.24$ & $5.0$ & $2.7$ & $\kappa$\\
FCC$_{\rm ee}$ & $0.24$+0.365 & $6.5$ & $1.3$ & $\kappa$\\
\hline\hline
\end{tabular}
\end{center}

\end{table}

\subsection{$M_s>M_h/2$}
\label{sec:singlethighmass}
The case of $M_s>M_h/2$ is more challenging. In this case, there is a universal modification of all $hxx$ couplings, see Eq. (\ref{eq:delzh}). The partial width into any SM final state, $\Gamma(h\to f)$, as well as the total Higgs width, $\Gamma_h$, change by the same factor, $(1+\delta g_h)^2$. Consequently, the branching ratios into the various final states are unchanged from the SM, but the production rates are modified:
\beq
\mu_i^f\equiv\frac{\sigma_i(pp\to h)\times{\rm BR}(h\to f)}{\left[\sigma_i(pp\to h)\times{\rm BR}(h\to f)\right]_{\rm SM}}=(1+\delta g_h)^2.
\eeq
Here $i$ is the Higgs production mode (ggF, VBF, {\it etc.}). If there is no $Z_2$ symmetry, then doublet-singlet mixing is allowed, {\it i.e.} $\sin2\theta\neq0$ (see Eq. (\ref{eq:hsmix})), and the dominant contribution to $\delta g_h$ in Eq. (\ref{eq:delzh}) is likely to come from the $(\cos\theta-1)$ term, and consequently
\beq
\mu_i^f\approx \cos^2\theta.
\eeq
If there is an unbroken $Z_2$ symmetry,
\beq
(\mu_i^f)^{(Z_2)}=1-\frac{|\lambda_{hs}|^2v^2}{8\pi^2}I_B(M_h^2;M_s^2,M_s^2).
\eeq
For $M_s^2\gg M_h^2$, we have
\beq
(\mu_i^f)^{(Z_2)}(M_s^2\gg M_h^2)\approx1+\frac{|\lambda_{hs}|^2}{48\pi^2}\frac{v^2}{M_s^2}.
\eeq

Under the assumption that the values of the signal strengths $\mu_i^f$ are the same for all production processes $i$ and decay channels $f$, a fit to the ATLAS and CMS data at $\sqrt{s}=7$ and 8 TeV, with $\mu$ as the parameter of interest, results in the best fit value \cite{Khachatryan:2016vau}:
\beq
\mu=1.09^{+0.11}_{-0.10}.
\eeq
With the full HL-LHC dataset, the LHC now projects, for the two best measured production rates, $1.6\%$ for the ggF channel and $3.1\%$ for the VBF channel. There is still a question of theory uncertainties on the ggF and VBF cross sections. One can argue that the theory precision on VBF will be $\lsim1\%$ while for ggF this is harder to estimate. In principle the above numbers already fold in acceptance uncertainties due to theory but not yet overall normalization uncertainties.

As concerns the ILC, Ref. \cite{Barklow:2017suo} (Table 6) estimates the accuracy on $\sigma_{Zh}$ to be $0.7\%$, while for CEPC and FCC$_{\rm ee}$ a precision of 0.5\% is expected. The accuracy on the measurement of $\sigma_{Zh}$ in various proposed $e^+e^-$ colliders is presented in Table \ref{tab:sigma}.

\begin{table}
\caption{Precision on the dominant Higgs production cross sections for the following $e^+e^-$ colliders: CLIC~\cite{Robson:2018zje}, CEPC~\cite{CEPCStudyGroup:2018ghi}, ILC~\cite{Barklow:2017suo} and FCC$_{\rm ee}$.}
\label{tab:sigma}
\begin{center}
\begin{tabular}{|c|c|c||c|} \hline\hline
\rule{0pt}{1.2em}%
Collider &  $\sqrt{s}$ [TeV] & ${\cal L}$ [ab$^{-1}$] & $\delta\sigma_{Zh}/\sigma_{Zh}$ [\%] \\[2pt] \hline\hline
\rule{0pt}{1.2em}%
CLIC & $0.38$ & $1.0$ & $1.3$\\
ILC & $0.25$ & $2.0$ & $0.7$ \\
CEPC & $0.25$ & $5.6$ & $0.5$ \\
FCC$_{\rm ee}$ & $0.24$ & $5.0$ & $0.5$ \\
\hline\hline
\end{tabular}
\end{center}
\end{table}

A particularly interesting issue in which the addition of a singlet scalar to the SM is relevant is the possibility that its coupling to the Higgs field makes the EWPT first order. Ref. \cite{Katz:2014bha} obtains that the lower bound on $|\lambda_{hs}|^2/M_s^2$ is such that
\beq
\mu_i^f-1 \gsim 0.6\%.
\eeq
We conclude that the ILC, CEPC and FCC$_{ee}$ may be able to lend
support or exclude this scenario as they reach a precision of 0.5-0.7\%. However, the sensitivity may only be at the level of one standard
deviation for all three colliders if the deviation is near its lower bound.

At CLIC, this scenario has been studied in detail~\cite{clicbsm} and it was estimated that, based on the Higgs coupling measurements, a constraint of
\beq
\mu_i^f-1\lsim 0.24\%\ \mbox{at 95\% C.L.}
\eeq
can be derived when including all stages, making it sensitive to probing the order of the EWPT.

The width measurements presented in Table~\ref{tab:width} are also sensitive to this scenario but the precision is inferior than that of the total cross sections.

\section{What is the solution of the flavor puzzle(s)?}
\label{sec:flavor}
There are several reasons that make the study of flavor physics via Higgs physics well motivated. First,
flavor physics raises three puzzles \cite{Nir:2013maa}:
\begin{itemize}
\item The Standard Model flavor puzzle: Why is there structure (smallness and hierarchy) in the charged fermion masses and the CKM mixing angles?
\item The neutrino flavor puzzle: Why, in contrast to the charged fermions, there seems to be no structure (neither hierarchy nor degeneracy nor smallness) in the neutrino-related flavor parameters?
\item The new physics flavor puzzle: If there is new physics at the TeV scale, what is the mechanism (alignment and/or degeneracy) that prevents it from significantly modifying the SM predictions for flavor changing neutral current processes?
\end{itemize}
Various models have been suggested to answer one or more of these
questions. The best hope to make further progress is by measuring new
flavor parameters (beyond the matrix elements of the CKM matrix). Measurements of the Yukawa couplings of $h$ provide such an opportunity.

Second, within the Standard Model (SM), flavor changing neutral current (FCNC) processes are suppressed by three factors:
\begin{itemize}
\item Loop suppression;
\item CKM suppression;
\item GIM suppression.
\end{itemize}
This unique situation allows the measurements of FCNC processes to probe new physics at very high scales. The search for Higgs-related flavor violating processes, such as $t\to ch$ or $h\to\tau\mu$, provides a new arena for FCNC measurements.

Within the SM, there is a clear prediction for the tree-level values of the Yukawa couplings:
\beq
Y_F=(\sqrt2/v)M_F\ \ \ (F=U,D,E).
\eeq
This relation between the Yukawa matrices and the mass matrices encompasses, in fact, three features:
\begin{itemize}
\item Proportionality: The diagonal Yukawa couplings ($y_f\equiv Y_{ff}$) are proportional to the corresponding masses. For example,
\beq\label{eq:pro}
y_\mu/y_\tau=m_\mu/m_\tau.
\eeq
\item Factor of proportionality: The factor that connects the mass to the diagonal Yukawa coupling is $\sqrt2/v$. For example,
\beq\label{eq:fac}
y_\tau=(\sqrt2/v)m_\tau.
\eeq
\item Diagonality: The off-diagonal Yukawa couplings vanish. For example,
\beq\label{eq:dia}
Y_{\mu\tau}=0.
\eeq
\end{itemize}
The three examples in Eqs. (\ref{eq:pro}), (\ref{eq:fac}) and (\ref{eq:dia}) demonstrate that, by measuring (or constraining) the rates of $h\to\tau\tau$, $h\to\mu\mu$ and $h\to\tau\mu$, we can test each of the three features predicted by the SM separately. In case that deviations from the SM predictions are established, the pattern of deviations will allow us to distinguish between various frameworks that aim to solve the flavor puzzles. This is demonstrated in Table \ref{tab:rxx}, where the predictions of various flavor models for the Yukawa couplings are given (adapted from \cite{Dery:2013rta}). In this Table, the scale $\Lambda$ is the scale that suppresses dimension-six terms in the SM effective-field-theory (SM-EFT) of the form
\beq
\frac{z_{ij}}{\Lambda^2}(\Phi^\dagger\Phi)\overline{L_i}\tilde\Phi E_j,
\eeq
where $L_i$ are the lepton doublet fields in the $(1,2)_{-1/2}$ representation, and $E_j$ are the lepton singlet fields in the $(1,1)_{-1}$ representation.

\begin{table}[t]
\caption{Predictions for di-lepton Higgs decays in various flavor models: SM, natural flavor conservation (NFC), minimal supersymmetric SM (MSSM), minimal flavor violation (MFV), Froggatt-Nielsen (FN) models and Giudice-Lebedev (GL). (See \cite{Dery:2013rta} for details.)}
\label{tab:rxx}
\begin{center}
\begin{tabular}{cccc} \hline\hline
\rule{0pt}{1.2em}%
 Model &\ $\frac{y_\tau}{\sqrt2 m_\tau/v}$   &  $\frac{y_{\mu}/y_{\tau}}{m_\mu/m_\tau}$\ &\ $\frac{{\rm BR}(h\to\mu\tau)}{{\rm BR}(h\to\tau\tau)}$\ \cr \hline
SM & $1$ & $1$ & $0$ \cr
NFC & $V_{h\ell}^*v/v_\ell$  & $1$ & $0$ \cr
MSSM & $\sin\alpha/\cos\beta$ & $1$ & $0$ \cr
MFV & $1+{\cal O}(v^2/\Lambda^2)$ & $1+{\cal O}(m_\tau^2/\Lambda^2)$ & $0$ \cr
FN & $1+{\cal O}(v^2/\Lambda^2)$ & $1+{\cal O}(v^2/\Lambda^2)$ & ${\cal O}(|U_{23}|^2v^4/\Lambda^4)$ \cr
GL & $3$ & $5/3$ & ${\cal O}[(25/9)(m_\mu^2/m_\tau^2)]$ \cr
\hline\hline
\end{tabular}
\end{center}
\end{table}

In Table \ref{tab:thq}, we present the predictions in various theoretical models for the flavor changing $t\to hq$ decays. The numerical estimates for the SM, 2HDM, MSSM and RS frameworks are taken from Ref. \cite{Agashe:2013hma} (Table 1-7). The parametric suppression factors in the SM-EFT subject to minimal flavor violation (MFV) or to Froggatt-Nielsen (FN) selection rules are taken from Refs. \cite{Dery:2013aba} and \cite{Dery:2014kxa}, respectively.

\begin{table}[t]
\caption{Predictions BR($t\to hq$) ($q=u,c$) in various models. (See \cite{Agashe:2013hma,Dery:2013aba,Dery:2014kxa} for details.)}
\label{tab:thq}
\begin{center}
\begin{tabular}{c|cccc|cc} \hline\hline
\rule{0pt}{1.2em}%
  & SM   & 2HDM & MSSM & RS & SM-EFT(MFV) & SM-EFT(FN) \cr \hline
$t\to hc$ & $3\times10^{-15}$ & $2\times10^{-3}$ & $\leq10^{-5}$ & $\leq10^{-4}$ & $y_b^4(V_{cb}V_{tb})^2(v/\Lambda)^4$ & $ |V_{cb}|^2(v/\Lambda)^4$ \cr
$t\to hu$ & $2\times10^{-17}$ & $6\times10^{-6}$ & $\leq10^{-5}$ & $-$ & $y_b^4(V_{ub}V_{tb})^2(v/\Lambda)^4$ & $|V_{ub}|^2(v/\Lambda)^4$ \cr
\hline\hline
\end{tabular}
\end{center}
\end{table}

In the following, the current measurements and expected accuracy of future measurements for the parameters relevant to flavor are discussed. For this purpose, we assume the future collider parameters presented in Section \ref{sec:exp}.

\begin{figure}[tb]
  \centering
  {\includegraphics[angle=270,width=0.65\textwidth]{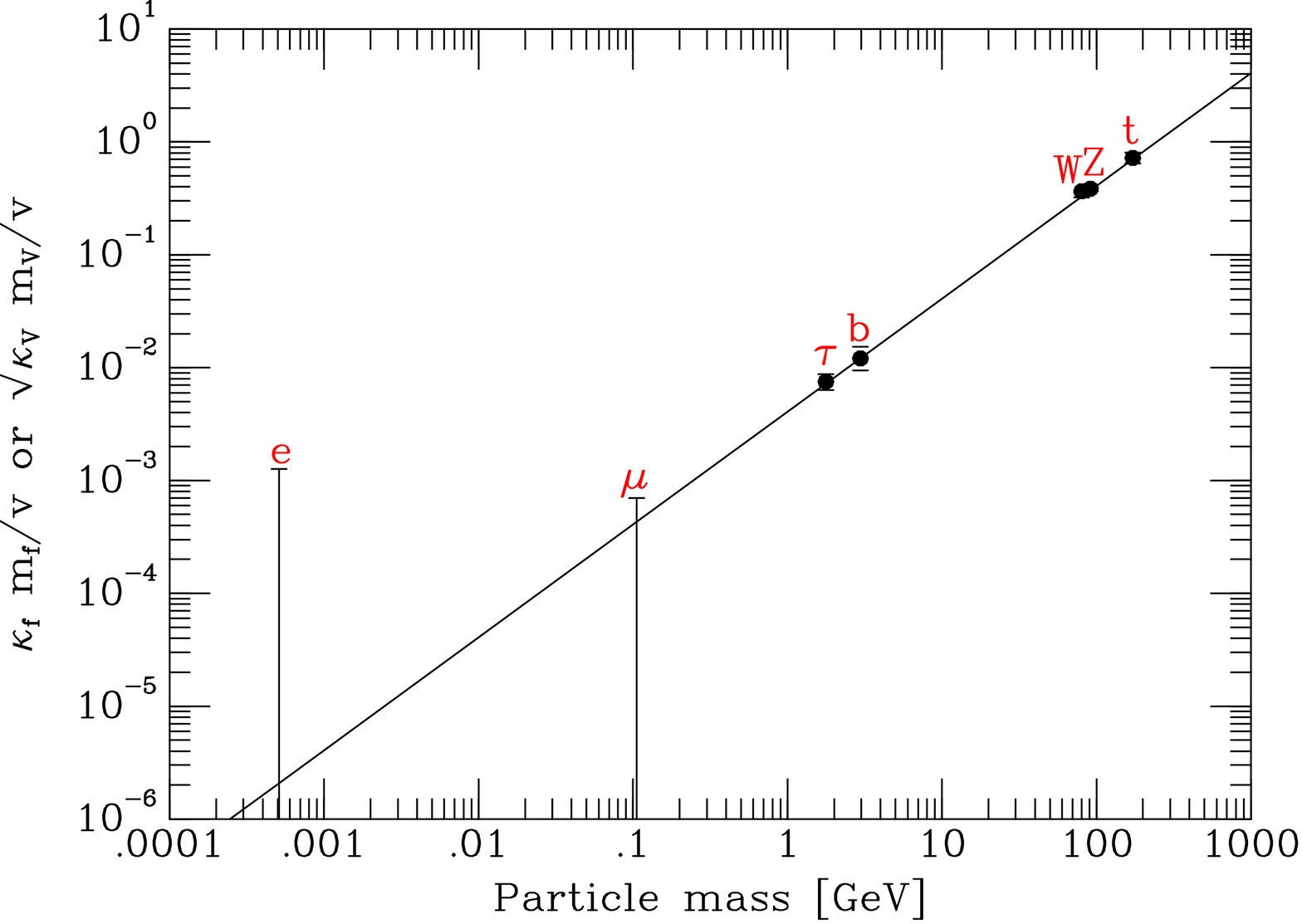}}
  \caption{The currently allowed experimental ranges for the Higgs couplings. For the Yukawa couplings, we present $\kappa_f m_f/v$. For the coupling to the electroweak vector bosons, we present $\sqrt{\kappa_V}m_V/v$.  The SM prediction is presented by the diagonal solid line. (Keith Ellis, private communication.)}
  \label{fig:yimi}
\end{figure}

The present status of the measured $y_f$ values is presented in Table
\ref{tab:yff}, and its compatibility with the SM prediction of
proportionality is presented in Fig. \ref{fig:yimi}. For the various
colliders, different methods are used to extract the couplings,
leading to some model-dependence at the level of $(0.5-1.0)$\%.  For the
third generation fermions the current LHC precision is about $(10-25)$\%. With
HL-LHC it will be improved to $(2-4)$\%. The other future machines will
improve the precision by another factor of $\sim 3$.

\begin{table}[t]
\caption{Experimental status and future projections of the diagonal Yukawa couplings $y_{f}$, and the accuracy estimated for future experiments in \%. The accuracy quoted for future experiments is based on combining those data with the HL-LHC, except for
CLIC and FCC240 where only the accuracy of the future collider is stated.
Upper bounds are given at 95\% confidence level. The first line in each row shows the CMS result and the second line the ATLAS result. In both cases, a BSM contribution is allowed, the $\gamma\gamma$ and $gg$ loop processes are treated with effective couplings, and the $Z\gamma$ process is resolved. For the $e^+e^-$ colliders, the same assumptions on the loops are made when using the so-called $\kappa$-framework. For LHC and HL-LHC~\cite{Cepeda:2019klc} , BR$_\textrm{BSM}=0$ is assumed.
For ILC an EFT fit is used to extract the values and the values are combined with HL-LHC~\cite{Bambade:2019fyw}. For all other colliders the $\kappa$-fit results are quoted from Refs,~\cite{clic,CEPCStudyGroup:2018ghi,Mangano:2018mur,lhec} and no combination with HL-LHC is made. All upper limits are given at 95\% C.L.. For the FCC-ee, an upper limit of 1.6 can be set on $y_{e}/y_e^{\rm SM}$ if one year of running at $\sqrt{s}\approx m_h$ is performed. When no value is available in the literature, a $-$ is shown.
}
\label{tab:yff}
\begin{center}
\begin{tabular}{c|c|ccccccccc} \hline\hline
\rule{0pt}{1.2em}%
 Observable &\ Current range & HL-LHC & ILC250 & ILC250+500 & CLIC380 & CLIC3000& CEPC & FCC240 & FCC365 & LHeC \\
                    &                          & \multicolumn{9}{c}{$\delta y/y$ (\%)}  \\ \hline
$y_{t}/y_t^{\rm SM}$ &  \begin{tabular}{@{}c@{}}$1.02^{+0.19}_{-0.15}$ \cite{CMSrun2hcomb}\\ $1.05^{+0.14}_{-0.13}$~\cite{ATLASrun2hcomb}\end{tabular} & 3.4& $-$ & $6.3$& $-$ & $2.9$ & $-$  & $-$  & $-$  & $-$  \cr\hline
$y_{b}/y_b^{\rm SM}$  & \begin{tabular}{@{}c@{}}$0.91^{+0.17}_{-0.16}$ \cite{CMSrun2hcomb}\\ $0.85^{+0.13}_{-0.14}$~\cite{ATLASrun2hcomb}\end{tabular}& 3.7 & $1.0$ & $0.60$ & $1.3$ & $0.2$ & 1.0& 1.4 & 0.67 & 1.1 \cr\hline
$y_{\tau}/y_\tau^{\rm SM}$ & \begin{tabular}{@{}c@{}}$0.93 \pm 0.13$ \cite{CMSrun2hcomb}\\ $0.95\pm 0.13$~\cite{ATLASrun2hcomb}\end{tabular} & 1.9 & $1.2$ & $0.77$ & $2.7$ & $0.9$& 1.2& 1.4 & 0.78& 1.3\cr\hline
$y_{c}/y_c^{\rm SM}$  & $<6.2$  \cite{Aaboud:2018fhh,PerezCharm} & $<220$ & $1.8$ & $1.2$ & $4.1$& $1.3$ & 1.9 & 1.8 & 1.2& 3.6\cr\hline
$y_{\mu}/y_\mu^{\rm SM}$   & \begin{tabular}{@{}c@{}}$0.72^{+0.50}_{-0.72}$\cite{CMSrun2hcomb}\\ $<1.63$~\cite{ATLASrun2hcomb}\end{tabular}  & 4.3 & $4.0$ & $3.8$& $-$ & $5.6$ & 5.0 & 9.6 & 3.4 & $-$  \cr\hline
$y_{e}/y_e^{\rm SM}$      & $<611$ \cite{Khachatryan:2014aep} & $-$ & $-$ & & $-$ & $-$ & $-$ & $-$ & $<1.6^{(+)}$ & $-$ \cr
\hline\hline
\end{tabular}
\end{center}
\end{table}

For the second (and first) generation fermions currently only upper limits are available. As concerns $y_\mu$, the HL-LHC will measure it to better than 5\%, and future $e^+e^-$ colliders are not expected to improve this further significantly. FCC$_{hh}$ will be able to measure the ratio ${\rm BR}(h\to\mu\mu)/BR(h\to 4\mu)$ with a precision of $1.3$\%. As concerns $y_c$, the HL-LHC expects to be sensitive to values 6-21 times larger than the SM value. Future colliders will probe it in earnest with a precision of (1-2)\%. At hadron colliders it is expected that very large anomalous values of the strange quark coupling can be probed on $\phi\gamma$ decays~\cite{Cepeda:2019klc}: for HL-LHC (FCC) the sensitivity is $-35<y_s/y_b^{\rm SM}<57$ ($-2<y_s/y_b^{\rm SM}<24$). At $e^+e^-$ colliders values of ${\cal O}(10)\times$SM rates might be in reach~\cite{Duarte-Campderros:2018ouv}.
As concerns the first generation, the
best hope for probing the SM value of $y_e$
directly is to run one of the circular $e^+e^-$ colliders at $\sqrt{s}\approx
m_h$ for several years. For the up and down quarks the projected
sensitivity of the HL-LHC is more than 1000 times the SM value, and is
only obtained indirectly.

The GL model (see Table~\ref{tab:rxx}) is already excluded by the current data presented in Table~\ref{tab:yff} (by both the $y_\tau$ measurement and the $y_\mu$ upper limit). For the FN model, the current constraints imply a lower bound on the scale of new physics, $\Lambda/\sqrt{z_{\tau\tau}} \gsim 600$~GeV. The expected precision of HL-LHC will increase the reach to $\Lambda/\sqrt{z_{\tau\tau}}\sim2$~TeV, while some of the future facilities can reach up to about $\Lambda/\sqrt{z_{\tau\tau}}\sim4$~TeV.

The present status of measurements and the future projections for off-diagonal branching ratios are given in Table \ref{tab:brij}. At present most upper limits are between 0.1\% and 1\%. With HL-LHC they will be improved by an order of magnitude to about $1\times 10^{-4}$.
For the $t\to qh$ decays, it can be expected that the sensitivity of
future $e^+e^-$ colliders is to branching ratios of order $10^{-5}$,
similar to other rare decay modes of the top
quark~\cite{Mangano:2018mur}, but no dedicated study is
available. With FCC-hh an improvement by another 1-2 orders of
magnitude is expected, depending on the systematic uncertainties that
will be obtained. These values are somewhat interesting in view of theories beyond the SM, see Table~\ref{tab:thq}.

\begin{table}[t]
\caption{Experimental status of measurements that depend on off-diagonal Yukawa couplings (assuming SM production rates). The LHC $t\to qh$ measurements are based on 36.1~fb$^{-1}$. For FCC$_{hh}$ the range quoted for $t\to qh$ reflects the size of the systematic uncertainty on the background of (0-5)\%.}
\label{tab:brij}
\begin{center}
\begin{tabular}{cccc} \hline\hline
\rule{0pt}{1.2em}%
 Observable &\ Upper bound   & HL-LHC & FCC-hh\ \cr \hline
${\rm BR}(t\to ch)$ & $1.1\times10^{-3}$ \cite{Aaboud:2018oqm} & $10^{-4}$~\cite{Azzi:2019yne} &  $(4.9-16)\times 10^{-6}$~\cite{Mangano:2018mur} \cr
${\rm BR}(t\to uh)$ & $1.2\times10^{-3}$ \cite{Aaboud:2018oqm} & $10^{-5}$~\cite{Azzi:2019yne} & $(2.5-28)\times 10^{-6}$~\cite{Mangano:2018mur}  \cr
${\rm BR}(h\to\tau\mu)$ & $2.5\times10^{-3}$ \cite{Aad:2016blu,Sirunyan:2017xzt} & $5\times 10^{-4}$~\cite{Cepeda:2019klc} & $-$\cr
${\rm BR}(h\to\tau e)$ & $6.1\times10^{-3}$ \cite{Aad:2016blu,Sirunyan:2017xzt} & $5\times 10^{-4}$~\cite{Atlas:2019qfx} & $-$\cr
${\rm BR}(h\to\mu e)$ & $3.4\times10^{-4}$ \cite{Khachatryan:2016rke} & $-$ & $-$ \cr
\hline\hline
\end{tabular}
\end{center}
\end{table}

Finally, let us mention that, while the third generation couplings are already constrained to be within about 10\% of the SM values, large deviations might still arise in measurements of the Yukawa couplings of the first two generations and of flavor changing decays. The experimental effort to search for these decays has thus the potential to make surprising discoveries, even if the sensitivity is far from the SM predictions.

\section{Additional Questions}
\label{sec:addque}
\subsection{Is $h$ alone?}
\label{sec:2hdm}
Within the SM, there is a single real scalar field, the Higgs boson $h_{\rm SM}$. It originates from a field $\Phi(1,2)_{1/2}$ that is a doublet of the $SU(2)$ gauge symmetry. In the unitary gauge, the Higgs field is given by
\beq
\Phi=\left(\begin{matrix} 0 \cr \frac{1}{\sqrt2}(v+h_{\rm SM})\cr\end{matrix}\right).
\eeq
The scalar $h$ discovered in the ATLAS and CMS experiments \cite{Aad:2012tfa,Chatrchyan:2012xdj} has measured properties that are all consistent, within present experimental accuracy, with the SM predictions for the Higgs boson \cite{Khachatryan:2016vau}. The question that is being asked is whether the observed Higgs boson is the SM Higgs boson.
In fact, one can separate this question to two:
\begin{enumerate}
\item Does $h$ originate from purely a doublet?
\item If there is more than a single doublet, is $h$ in exactly the direction of $v$?
\end{enumerate}
As concerns the first question, an example of a scenario where $h$ does not originate from purely a doublet is that of doublet-singlet mixing, discussed in Section \ref{sec:singlet}. As concerns the second question, we briefly discuss in this subsection the scenario of two Higgs doublet models (2HDM).

In a 2HDM with Natural Flavor Conservation (NFC), there are three well defined bases:
\begin{enumerate}
\item The NFC basis $(\Phi_1,\Phi_2)$, where each fermionic sector couples to either $\Phi_1$ or $\Phi_2$, but not to both.
\item The Higgs basis $(\Phi_v,\Phi_A)$, where $\langle\Phi_v\rangle=v/\sqrt2$ and $\langle\Phi_A\rangle=0$.
\item The mass basis $(\Phi_h,\Phi_H)$, which is the mass basis for the CP-even neutral scalars $h$ and $H$.
\end{enumerate}
The rotation angle from the  $(\Phi_1,\Phi_2)$ to the $(\Phi_v,\Phi_A)$ [$(\Phi_h,\Phi_H)$] basis is denoted $\beta$ [$\alpha$]. We define $\kappa_x$ via
\beq
\kappa_x\equiv\frac{g_{hxx}}{g_{hxx}^{\rm SM}}.
\eeq
In all 2HDMs we have ($V=W,Z$)
\beqa
\kappa_V&=&\sin(\beta-\alpha)\simeq1-\frac12\cos^2(\beta-\alpha),\\
\kappa_h&\simeq&1-\frac{2m_A^2}{m_h^2}\cos^2(\beta-\alpha).\no
\eeqa
Note that for $m_A^2\gg m_h^2$, we have $(m_A^2/m_h^2)\cos^2(\beta-\alpha)\simeq\cos(\beta-\alpha)$.
In all NFC models, where by definition $\Phi_1$ is the doublet that does not couple to the up sector,
\beqa
\kappa_t&=&\sin(\beta-\alpha)+\frac{\cos(\beta-\alpha)}{\tan\beta},\\
\kappa_b&=&\sin(\beta-\alpha)-\cos(\beta-\alpha)\tan\beta\ \ {\rm or}\ \ \kappa_t,\no\\
\kappa_\tau&=&\sin(\beta-\alpha)-\cos(\beta-\alpha)\tan\beta\ \ {\rm or}\ \ \kappa_t,\no
\eeqa
where the two options on the RHS of $\kappa_{b}$ correspond to the cases where $b$ couples to $\Phi_1$ or $\Phi_2$, respectively, and similarly for $\kappa_\tau$. The largest deviations in these models arise in the down or charged lepton sector, if either or both couple to the doublet to which the up sector does not couple. If all fermions couple to $\Phi_2$, then the largest deviation is expected in $\lambda_h$ \cite{Efrati:2014uta}. At tree level, the results for the MSSM are the same as those of a type II 2HDM.

We conclude that the most promising precision measurements of Higgs decays for obtaining hints of 2HDM are those of $h\to b\bar b$ and $h\to\tau^+\tau^-$. The future projections of measuring $y_b$ and $y_\tau$ in future experiments can be read off Table \ref{tab:yff}. An even more powerful probe of 2HDM is the combination of these two couplings with a measurement of $\kappa_V$, see Table \ref{tab:sigma}.
With $\kappa_Z=\sin(\beta-\alpha)$ measured very accurately, an accuracy of ${\cal O}(1\%)$ on $y_\tau$ will essentially constrain $\cos(\beta-\alpha)\tan\beta$ at that level in NFC models where the $\tau$ couples to $\Phi_1$ only.

\subsection{Is $h$ elementary?}
\label{sec:CH}
The question of Higgs compositeness is interesting in its own sake, but also in the context of mechanisms to explain the smallness of $m_h^2$ compared to $m_{\rm Pl}^2$. Various models have been proposed where the Higgs is a pseudo-Goldstone boson which emerges from a strongly interesting sector \cite{ArkaniHamed:2002qy,Contino:2003ve,Agashe:2004rs}. An effective field theory framework for the strongly interacting light Higgs (SILH) incorporates the basic features of a large class of these models \cite{Giudice:2007fh}. The low energy effective SILH Lagrangian depends on essentially two parameters: $m_\rho$, the mass scales of new resonances emerging from the strongly interacting sector, and $g_\rho$, their coupling. It is useful to define the dimensionless combination
\beq
\xi=g_\rho^2 v^2/m_\rho^2.
\eeq
The terms in the SILH Lagrangian that are relevant to our purposes are given by
\beq
{\cal L}_{\rm SILH}=\frac{c_H\xi}{2v^2}\partial^\mu(\phi^\dagger\phi)\partial_\mu(\phi^\dagger\phi)
+\frac{c_y\xi y_f}{v^2}\phi^\dagger\phi\overline{f_L}\phi f_R+{\rm h.c.}.
\eeq
Naive dimensional analysis (NDA) suggests that $c_H$ and $c_y$ are ${\cal O}(1)$. The theoretical framework only allows the parameters to be in the range
\beq
1\leq g_\rho\leq4\pi,\ \ \  \xi\leq1.
\eeq
Electroweak precision measurements give stronger constraints. The new physics contribution to $\hat S$ and $\hat T$ can be parameterized as follows \cite{Thamm:2015zwa}:
\beqa
\Delta\hat S&=&\frac{g^2}{96\pi^2}\xi\left[\log\left(\frac{\Lambda}{m_h}\right)+6\alpha\right]+\frac{m_W^2}{m_\rho^2},\no\\
\Delta\hat T&=&-\frac{g^{\prime2}}{32\pi^2}\xi\left[\log\left(\frac{\Lambda}{m_h}\right)-6\beta\frac{y_t^2}{g^{\prime2}}\right],
\eeqa
where $\Lambda=4\pi m_\rho/g_\rho$ and the coefficients $\alpha,\beta={\cal O}(1)$ and could have either sign. The resulting constraints read \cite{Thamm:2015zwa}
\beq
\xi\leq0.15,\ \ \ m_\rho\geq3\ TeV
\eeq
when marginalizing over $\alpha$ and $\beta$ (or $\xi\leq0.025$ and $m_\rho\geq4\ TeV$ for $\alpha=\beta=0$).

The modifications of the Higgs couplings are given by
\beqa
\Delta g_V/g_V^{\rm SM}&=&-c_H\xi/2,\no\\
\Delta g_f/g_f^{\rm SM}&=&-c_H\xi/2-c_y\xi,\no\\
\Delta g_g/g_g^{\rm SM}&=&-c_H\xi/2-c_y\xi,\no\\
\Delta g_\gamma/g_\gamma^{\rm SM}&=&-c_H\xi/2+0.3c_y\xi.
\eeqa
Thus, the maximal possible deviations are \cite{Gupta:2012mi}
\beqa
\Delta g_V/g_V^{\rm SM}&\approx&-0.08 c_H,\no\\
\Delta g_f/g_f^{\rm SM}&=&\Delta g_g/g_g^{\rm SM}\approx -0.08-0.15\frac{c_y}{c_H}\no\\
\Delta g_\gamma/g_\gamma^{\rm SM}&\approx&-0.08+0.05\frac{c_y}{c_H}.
\eeqa
E.g. for $c_H=c_y=1$ the maximal deviations are $-0.08$, $-0.20$ and $-0.03$ for the $g_V$, $g_{f,g}$ and $g_\gamma$ couplings, respectively. The estimates of the reach on $c_H\xi$ of Higgs-related measurements for various colliders are collected in Table \ref{tab:ch}. Comparing this to the maximal possible deviations given above, it is clear that future colliders will probe this aspect in a very interesting regime. Currently the precision obtained by ATLAS and CMS on $\kappa_Z$ and $\kappa_W$ is about 10\% each~\cite{CMSrun2hcomb,ATLASrun2hcomb}, resulting in a sensitivity of about 20\% on $c_H\xi$. Ref. \cite{Liu:2018qtb} employs a more general framework, based on nonlinear shift symmetries acting on $h$ and assuming custodial symmetry in the strong sector, and quotes the presently allowed 68\% CL range as
\beq
\xi=-0.041^{+0.090}_{-0.094}.
\eeq

\begin{table}[h!]
\caption{Summary of the reach on $\xi$ for various collider
  options, based on Ref. \cite{Thamm:2015zwa} but updated using the
  recent projections on the $\kappa_V$ sensitivity. See discussion in Section \ref{sec:CH}.}
\label{tab:ch}
\begin{center}
\begin{tabular}{ccccc} \hline\hline
\rule{0pt}{1.0em}%
Collider & Energy & Luminosity & $\Delta g_V/g_V$  (\%) & $(c_H\xi)_{max}\ [1\sigma]$  \\[2pt] \hline\hline
\rule{0pt}{1.0em}%
HL-LHC & 14 TeV & 3 ab$^{-1}$ & 1.3 & $0.026$ \\ \hline
ILC & 250 GeV & 2 ab$^{-1}$ & 0.56 &  0.011\\
    & +500 GeV & 4 ab$^{-1}$ & 0.38 & 0.008  \\ \hline
CLIC & 380 GeV & 1 ab$^{-1}$ & 0.6 & $0.012$  \\
 & +1.4 TeV & 2.5 ab$^{-1}$ & 0.6 & $0.012$ \\
    & +3.0 TeV & 5 ab$^{-1}$ & 0.6 & $0.012$  \\ \hline
FCC$_{ee}$ & 240 GeV & $2\times 5.0$ ab$^{-1}$ & 0.2& 0.004 \\
    & +350 GeV & 2.6 ab$^{-1}$ & 0.17 & $0.003$  \\\hline
CEPC & 240 GeV & $2\times 5.6$ ab$^{-1}$ & 0.25 &  0.005\\
\hline\hline
\end{tabular}
\end{center}
\end{table}

Ref. \cite{Thamm:2015zwa} compared the sensitivity of the Higgs related measurements and the direct resonance searches at (and beyond) the LHC within the framework of composite Higgs models. They reach the conclusion that the Higgs related measurements will be more sensitive for large $g_\rho$. Specifically, the indirect measurements explore novel territory for $g_\rho\gsim4.5$.

\subsection{What keeps $m_h^2\ll m_{\rm Pl}^2$?}
If the SM is only a low energy effective theory, and at some high scale $\Lambda_{\rm NP}$ there are new degrees of freedom which couple to the Higgs boson, then the mass-squared of the Higgs boson get corrections of ${\cal O}(\Lambda_{\rm NP}^2)$, and a fine-tuned cancelation with the bare $\mu^2$ term is needed to keep the Higgs mass light. Possible solutions to this fine-tuning problem can be classified to symmetry-related mechanisms, and dynamical scenarios.

The two most intensively studied frameworks where the solution is symmetry related are composite Higgs and Supersymmetry. We discussed the composite Higgs framework in Section \ref{sec:CH}.

As concerns the Higgs couplings in the minimal supersymmetric SM (MSSM), the situation is similar to the 2HDM discussed in Section \ref{sec:2hdm} with NFC type II, and with some additional constraints arising from relations to processes involving the supersymmetric partners of the SM degrees of freedom. Ref.~\cite{Gupta:2012mi} obtained the maximal possible deviations of the $h$ couplings, assuming that the scalar degrees of freedom of the second Higgs doublet are too heavy to be directly observed at the LHC. They obtain
\beqa
1-\kappa_V&\lsim&0.01,\no\\
1-\kappa_t&\lsim&0.03\no\\
1-\kappa_b&\lsim&0.1-1.
\eeqa

A dynamical solution to the fine tuning problem is provided by relaxion models \cite{Graham:2015cka}. The Higgs mass depends on a time-dependent VEV of a scalar field which rolls until it stops at a value much smaller than the cut-off scale of the theory. Unlike models of composite Higgs or Supersymmetry, where many new degrees of freedom are predicted at a scale not much higher than the electroweak symmetry breaking scale, in this framework there is a single new scalar, SM-singlet, degree of freedom, the relaxion $\phi_R$. Much of the relevant phenomenology comes from the generic situation where the relaxion and the Higgs mix \cite{Flacke:2016szy}. Thus, the relaxion inherits the Higgs couplings, suppressed by the small $\phi_R-h$ mixing angle. This scenario was discussed in Section \ref{sec:singlet}, particularly in the case of $M_s<M_h/2$.

\subsection{Was the electroweak phase transition (strongly) first order?}
At a temperature of order 100 GeV, the Universe went through a transition from a high temperature symmetric phase ($\langle\Phi\rangle=0$) to a state with broken electroweak symmetry ($\langle\Phi\rangle=v/\sqrt2\neq0$). If the SM is a good low energy effective theory up to the TeV scale, then the electroweak phase transition (EWPT) is a crossover transition. If, on the other hand, the EWPT was first-order, then electroweak baryogenesis was possible. In fact, successful EWBG requires that the phase transition is strongly first order,
\beq
\frac{v(T_c)}{T_c}\gsim1,
\eeq
where $T_c$ is the temperature at which the phase transition takes place. This requirement usually implies new degrees of freedom that are not much heavier than the EW scale and which couple to the Higgs field with a coupling that is not small.

The new physics that can make a first-order EWPT can be broadly classified to four classes \cite{Chung:2012vg}, distinguished by the physics that is responsible for generating the barrier between the EW-symmetric and EW-breaking minima:
\begin{itemize}
\item Thermally driven. New bosonic degrees of freedom generate a finite-temperature term in the effective potential of the form $-T(h^2)^{3/2}$.
\item Tree-level renormalizable operators. Additional scalars, typically $SU(2)_L$-singlets or doublets, mix with the Higgs boson and generate an effective, temperature-independent, $\phi^3$ term.
\item Tree level non-renormalizable operators. The most intensively studied example is a dimension-six term $(\Phi^\dagger\Phi)^3$ (with a negative $\lambda$ for the $\lambda(\Phi^\dagger\Phi)^2$ term).
\item Loop corrections. The most intensively studied example is an effective term of the form $\kappa h^4\ln(h^2/M^2)$ in the Higgs potential.
\end{itemize}

Loops involving scalar particles induce a cubic term in the effective potential at high temperature that can modify the EWPT into a first-order one (see, for example, Ref.~\cite{Katz:2014bha}). Consider then a new scalar $S$, with coupling to the Higgs field $\Phi$:
\beq\label{eq:ksp}
V_{S\Phi}=\kappa|S|^2|\Phi|^2.
\eeq
To make the phase transition first-order, we need $\kappa={\cal O}(1)$ and $m_S\lsim400$ GeV. This term modifies the following Higgs observables:
\begin{enumerate}
\item If $S$ is colored, the $hgg$ coupling is modified. Thus, at hadron colliders, a measurement of $\mu_{\rm ggF}$ is best suited to probe this scenario, which is expected to have a precision of about 2\% at HL-LHC. At lepton colliders, the best observable is BR($h\to gg$) which can be probed with a precision of about 1.5-2.5\%.
\item If $S$ is electromagnetically charged, the $h\gamma\gamma$ coupling is modified. Thus, a measurement of BR($h\to {\gamma\gamma}$) is best suited to probe this scenario. At HL-LHC a precision of about 2.6\% is expected, and lepton colliders are not expected to improve this significantly. A substantial improvement is expected from FCC$_{hh}$ with about 0.5\%.
\item Independently of the quantum numbers of $S$, the $hZZ$ coupling will be modified, since $S$ renormalizes the Higgs wavefunction \cite{Englert:2013tya,Craig:2013xia}. The case where $S$ is neither colored nor electromagnetically charged was discussed in Section \ref{sec:singlet}.
\end{enumerate}

Models of strongly first-order EWPT due to renormalizable tree-level operators will be further tested by Higgs measurements of the invisible Higgs decays or by the bounds on the mixing with singlet scalars, discussed in Section \ref{sec:singlet}, and by various other measurements (see {\it e.g.} \cite{Profumo:2014opa,Kotwal:2016tex}). For models of strongly first-order EWPT due to non-renormalizable tree-level operators,
\beq
V=\mu^2(\Phi^\dagger\Phi)+\lambda(\Phi^\dagger\Phi)^2+\frac{\rho}{\Lambda^2}(\Phi^\dagger\Phi)^3,
\eeq
the main consequence is a modification of the relation between the Higgs mass, the VEV and the trilinear Higgs self-coupling:
\beq
\lambda_{hhh}=3\frac{m_h^2}{v}+6\rho\frac{v^3}{\Lambda^2}.
\eeq
The required size of $\rho v^2/\Lambda^2$ is such that deviation of ${\cal O}(1)$ in $\lambda_{hhh}$ are predicted. Table ~\ref{tab:lhhh} summarizes the expected sensitivity of various future colliders.

\begin{table}[h!]
\caption{Summary of the reach on $\lambda_{hhh}$ for various collider options. Given are the collider, the center-of-mass energy, the integrated luminosity and the precision that is expected to be achieved on $\lambda_{hhh}$. For all colliders, except FCC$_{ee}$, it is based on di-Higgs measurements. For FCC$_{ee}$ it is based on the $\sqrt{s}$ dependence of higher order corrections to the $Zh$ cross section. In all cases, the luminosity quoted is that of the sum of the experiments. The values correspond to the luminosities given in Sec.~\ref{sec:exp}}
\label{tab:lhhh}
\begin{center}
\begin{tabular}{lcr} \hline\hline
\rule{0pt}{1.0em}%
Collider & $\sqrt{s}$ & $\frac{\Delta \lambda_{hhh}}{\lambda_{hhh}}$  (\%) \\[2pt] \hline\hline
\rule{0pt}{1.0em}%
HL-LHC & 14 TeV & $\pm 50$ \\ \hline
ILC & 500 GeV & $\pm 17$ \\ \hline
CLIC & 3.0 TeV & $^{+11}_{-7}$   \\ \hline
FCC$_{ee}$ & 240+365 GeV & $\pm 40$ \\\hline
HE-LHC & 27 TeV & $\pm$(10-20) \\\hline
FCC$_{hh}$ & 100 TeV & $\pm 5$ \\
\hline\hline
\end{tabular}
\end{center}
\end{table}

\subsection{Do CP violating $h$ interactions generate the baryon asymmetry?}
Complex Yukawa couplings could play a CP violating role in electroweak baryogenesis, see {\it e.g.} Refs. \cite{Shu:2013uua,Kobakhidze:2015xlz,Chiang:2016vgf,Guo:2016ixx}. Various suggestions of how to measure such CP violating Higgs interactions have been made, in particular for $h\tau\tau$ and $htt$ couplings, see {\it e.g.} Refs. \cite{Harnik:2013aja,Demartin:2014fia,Buckley:2015vsa,AmorDosSantos:2017ayi,
Huang:2015izx,Hayreter:2016kyv,Goncalves:2018agy,Bernlochner:2018opw}.

The ATLAS collaboration \cite{atlas:cpv} has recently estimated the accuracy of measuring the CP violating mixing angle $\phi_\tau$ defined via
\beq
{\cal L}_{h\tau\tau}=g_{h\tau\tau}[\cos(\phi_\tau)\bar\tau\tau+\sin(\phi_\tau)\bar\tau i\gamma_5\tau]h.
\eeq
With 3 ab$^{-1}$ at $\sqrt{s}=14$ TeV, $\phi_\tau$ can be measured with a statistical precision between $\pm18^o$ and $\pm33^o$, depending on the energy resolution of the $\pi^0$ achieved in the reconstruction at high pile-up.

\section{Conclusions}
\label{sec:con}
A list of interesting theoretical questions, and a partial list of observables that are most relevant to making progress on these questions, are summarized in Table \ref{tab:QO}.
\begin{table}[t]
\caption{A list of interesting theoretical questions, and a partial list of observables that are most relevant to making progress on these questions. $\kappa_{3}\ (\kappa_{\ell})$ stands for third (first or second) generation couplings. $\mu_{4f}$ stands for the processes $h\to SS\to\overline{f_1}f_1\overline{f_2}f_2$. For more details, see the text.}
\label{tab:QO}
\begin{center}
\begin{tabular}{c|cccccccccccc} \hline\hline
\rule{0pt}{1.2em}%
Question &  $\kappa_{V}$ & $\kappa_{3}$ & $\kappa_{g}$ & $\kappa_{\gamma}$ & $\lambda_{hhh}$ & $\sigma_{hZ}$ & BR$_{\rm inv}$ & BR$_{\rm und}$ & $\kappa_{\ell}$ & $\mu_{4f}$ & BR$_{\tau\mu}$ & $\Gamma_h$ \\[2pt] \hline\hline
\rule{0pt}{1.2em}%
Is $h$ Alone? & $+$ & $+$ &&& $+$ & $+$ &&&& $+$ && $+$ \\
Is $h$ elementary? & $+$& $+$ & $+$ & $+$ &&$+$&&&&&& \\
Why $m_h^2\ll m_{\rm Pl}^2$? & $+$ & $+$ &&&&& $+$ & $+$ && $+$ && $+$ \\
1st order EWPT? &&& $+$ & $+$ & $+$ & $+$ &&&& $+$ && \\
CPV? && $+$(CP) &&&&&&&&&& \\
Light singlets? &&&&&&& $+$ & $+$ & $+$ & $+$ && $+$ \\
Flavor puzzles? &&$+$&&&&&&& $+$ && $+$ &
\\
\hline\hline
\end{tabular}
\end{center}
\end{table}

Our review of what can be learned from Higgs precision measurements led us to three important conclusions, independent of the particular experiments that will perform these measurements:
\begin{itemize}
\item The Higgs program, which is a guaranteed program of any future collider, touches some of the most significant open questions in particle physics and cosmology. In particular, this program may lead to progress in two puzzles of particle cosmology: dark matter and the baryon asymmetry.
\item No Higgs related measurement will go unnoticed. Each of the production modes and each of the decay modes carries in it information that is relevant to important questions. Higgs decays to two bosons, two fermions, invisible modes, four fermions, as well as Higgs production by gluon-gluon fusion, $t\bar th$, and di-Higgs production, all of these (and other measurements) are highly valuable for better understanding of particle physics and cosmology.
\item For many of the interesting observables, the proposed future colliders significantly extend the sensitivity beyond that reachable by the HL-LHC alone.
\end{itemize}

\subsection*{Acknowledgements}
We thank the members of the Scientific Policy Committee (SPC) of CERN for initiating discussions that led to this work, and for comments on the manuscript. We thank the participants of the workshop ``Voyages Beyond the Standard Model II," and Gilad Perez and Georg Weiglein for useful discussions and comments on the manuscript.
BH acknowledges support by the Deutsche Forschungsgemeinschaft (DFG, German Research Foundation) under Germany's Excellence Strategy -- EXC 2121 "Quantum Universe" -- 390833306.
YN is the Amos de-Shalit chair of theoretical physics, and is supported by grants from the Israel Science Foundation (grant number 394/16), the United States-Israel Binational Science Foundation (BSF), Jerusalem, Israel (grant number 2014230), and the I-CORE program of the Planning and Budgeting Committee and the Israel Science Foundation (grant number 1937/12).


\end{document}